# Photo-Thermal Actuation of Hybrid Microgels with Dual Laser Optical Tweezers

Se-Hyeong Jung[a]†, Chi Zhang[b]†, Nick Stauffer[a], Frank Scheffold[b], Lucio Isa[a]*

| | |
|---|---|
| [a] | Dr. Se-Hyeong Jung, Nick Stauffer, Prof. Dr. Lucio Isa |
| | Department of Materials |
| | ETH Zurich |
| | Vladimir-Prelog-Weg 1-5/10, 8093 Zurich, Switzerland |
| | E-mail: lucio.isa@mat.ethz.ch |
| [b] | Dr. Chi Zhang, Prof. Dr. Frank Scheffold |
| | Department of Physics |
| | Université de Fribourg |
| | Ch. Du Musée 3, 1700 Fribourg, Switzerland |

[†] These authors contributed equally to this work.

Supporting information for this article is given via a link at the end of the document.

**Abstract:** Soft actuators that respond to external stimuli play a fundamental role in microscale robotics, active matter, and bio-inspired systems. Among these actuators, photo-thermal hybrid microgels (HMGs) containing plasmonic nanoparticles enable rapid, spatially controlled actuation via localized heating. Understanding their dynamic behavior at the single-particle level is crucial for optimizing performance. However, traditional bulk characterization methods such as dynamic light scattering (DLS), provide only ensemble-averaged data, thereby limiting analytical insights. Here, we introduce a dual-laser optical tweezers approach for real-time, single-particle analysis of HMGs under controlled light exposure. Combining direct imaging and mean-square displacement (MSD) analysis, our method quantifies the precise laser power required for actuation and accurately tracks the particle size. We benchmark our results against an existing dual-laser DLS, demonstrating comparable precision while offering the unique advantage of single-actuator resolution. Thus, our method provides as a robust platform for precise optimization of programmable actuators with applications in soft robotics, microswimmers, and biomedical applications.

*Introduction*

Soft actuators enable programmable shape transformations and mechanical responses to external stimuli, making them fundamental components of adaptive materials.[1-4] Their high deformability and responsiveness provide a key advantage over rigid actuators, particularly at the microscale, where low energy consumption and precise adaptation to stimuli such as light, temperature, pH, and magnetic fields are crucial.[5-7] These properties have been leveraged in diverse microscale applications, including thermo-responsive microvalves for fluidic control, light-driven microswimmers for targeted transport, and soft micromechanical oscillators for sensing.[8-12]

Among microscale soft actuators, microgels (MGs) stand out due to their reversible size change in response to a broad range of external stimuli. Microgels are defined as "particles of gel of any shape with an equivalent diameter of approximately 0.1 to 100 µm" according to IUPAC.[13] These crosslinked polymer networks undergo swelling/deswelling as a function of environmental conditions, such as temperature, ionic strength or pH. Adaptation to different stimuli therefore allows the dynamic tuning of size, shape, porosity, mechanical properties, and chemical composition of individual particles, [14-21] leading to self-regulated, programmable actuation without complex external control systems.[22-25] A prominent example are poly(N-isopropylacrylamide) (PNIPAM)-based microgels, which undergo a sharp, reversible phase transition from a swollen, hydrophilic state to a collapsed, hydrophobic state above their lower critical solution temperature (LCST, ~32°C).[16, 20, 26, 27] This sharp transition makes PNIPAM microgels highly effective for precise shape transformations in microscale soft actuation.

Although PNIPAM-based microgels exhibit efficient and reversible actuation, bulk thermal activation via global temperature modulations limits response speed, spatial precision, and independent control of individual actuators. To address this challenges, researchers have developed photo-thermal hybrid microgels (HMGs) by incorporating plasmonic nanoparticles capable of efficiently absorbing light and generating localized heat via surface plasmon resonance.[28-31] Among these, gold nanoparticles (AuNP) are widely used due to their strong plasmonic response at 532 nm, enabling rapid heating and selective microgel collapse.[32] This light-driven actuation strategy has unlocked new possibilities in microscale systems, such as photo-controlled Pickering emulsion[29], nanomechanical oscillators[33] and plasmonic self-assembled materials[34].

To fully exploit photo-thermal HMGs in soft actuation, precise characterization of their responses at the single-particle level is essential. Traditional methods such as UV-VIS spectroscopy track bulk optical shifts during irradiation,[35] while temperature-controlled studies monitor ensemble shrinkage behavior.[29] To this end, an advanced dual-laser DLS technique[32] has been proposed to provide in situ size measurements under laser irradiation. Here, one laser is used to excite the metallic nanoparticles and cause local heating, while the other one probes the size changes. However, bulk DLS analysis is fundamentally limited by averaging effects, global heating artifacts dependent on particle concentration, and the assumption of a uniform light absorption and heat conversion for all HMGs. Thus, heterogeneous particle behavior and localized responses remain unresolved, particularly in mixed or polydisperse actuator systems.

To overcome these limitations, we introduce a dual-laser optical tweezers method that enables real-time analysis of single-particle HMG responsiveness under controlled light exposure. By directly visualizing microgel size changes through image tracking and extracting dynamic properties through MSD analysis, our approach precisely quantifies the laser power required for particle collapse, resolving variations that ensemble techniques cannot detect. To evaluate the consistency of our dual-laser optical tweezers approach, we benchmark it via dual-laser DLS. We extend the measurements achieved in previous studies to a larger scattering angle of 90°[32] and by accounting for the drift induced by laser radiation pressure, we achieve improved accuracy. Direct comparisons confirmed that both methods yield similar results, validating our optical tweezers data. However, optical tweezers provide additional crucial advantages, enabling real-time visualization and precise characterization under conditions closely resembling practical applications. Beyond HMGs, this powerful single-particle analytical method can be broadly applied to other microscale systems such as colloidal actuators, light-driven microswimmers, and reconfigurable soft materials, opening new possibilities in responsive material design.

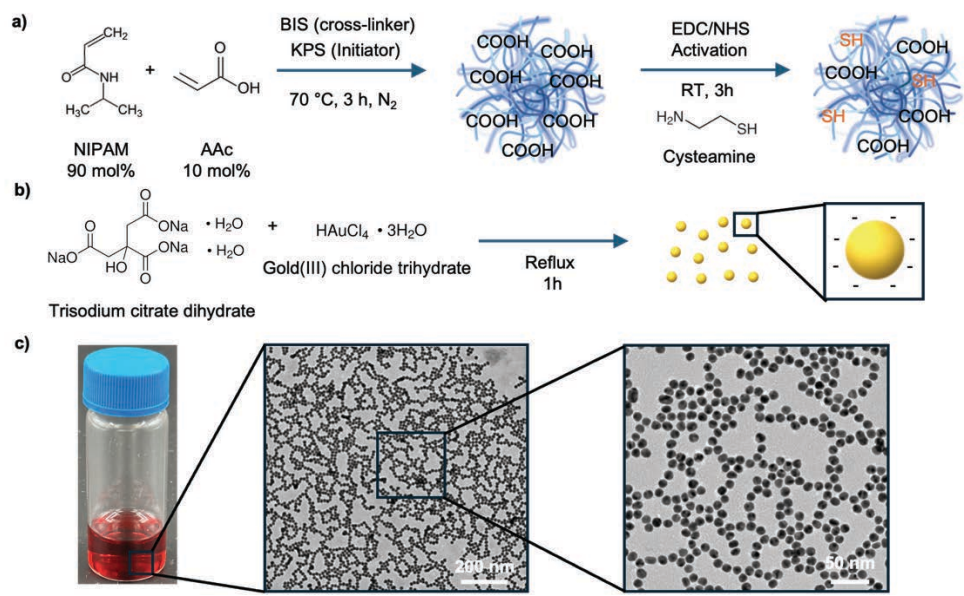

**Figure 1.** Schematic illustration of the synthesis and modification steps for microgels (a) and AuNPs (b). (c) shows an image of AuNPs dispersion in water, along with TEM images of the synthesized AuNPs in dry state.

*Results and discussion*

A narrow plasmonic absorption band and large volumetric response are central to the precise actuation of HMGs. We therefore first systematically optimized our particle systems. Initially, PNIPAM-co-poly(acrylic acid) (PAAc) MGs with COOH groups were synthesized via precipitation polymerization (**Fig. 1a**). The COOH groups in the MGs, which confer negative charges at pH around 6 (miliQ water)[36], ensure colloidal stability. The characterization of these MGs showed a hydrodynamic radius ($R_h$) of 857 ± 13.28 nm and a zeta potential of -14.5 ± 0.28 mV, conforming the incorporation of COOH groups.[37, 38] The COOH groups were subsequently substituted with cysteamine through EDC/NHS coupling for activation.[39, 40] The MG's zeta potential slightly increases and the size decreases with higher thiol (SH) content (from 12.5 % to 50 %), corresponding to a decrease in the number of COOH groups (**Tab. S1**). This reduction in surface charge leads to lower size due to weaker electrostatic interactions and osmotic pressure change.[37, 38] Next, we synthesized AuNPs by reducing $Au^{3+}$ salts with citrate according to the Turkevich method[41-43], as shown in **Figure 1b**. The resulting monodisperse AuNPs exhibited the characteristic red colour (**Fig. 1c**), indicating a stable dispersion, with a zeta potential of -33 ± 2.36 mV. TEM images revealed a uniform particle size of 12 ± 0.21 nm (**Fig. 1c** and **Fig. S1**), compatible with DLS measurements indicating a $R_h$ of 18 ± 0.04 nm, accounting for the core and surrounding solvation layer, adsorbed ions, or stabilizing agents.[44] (The DLS and zeta potential measurements of the MGs and AuNPs are summarized in **Tab. S1**).

Next, we prepared HMGs via hetero coagulation by mixing microgels and AuNPs, taking advantage of the strong binding affinity of SH groups toward metallic nanoparticles (**Fig. 2a**).[45-47] The ratio between COOH and SH groups plays a crucial role in controlling aggregation behavior. Elemental analysis confirms that the sulfur (S) content increases with higher cysteamine concentrations (**Tab. S2**). TEM images (**Fig. 2b-d**) reveal that increasing SH content from 12.5 % to 25 % and 50 % leads to AuNP clustering within the microgels. The 12.5 % modification enables high loading of the AuNPs (> 1000 AuNPs per microgel, **Fig. S2**) while preventing aggregation and is thus chosen for the subsequent analytical experiments. This trend is further reflected in bulk solution color changes, where higher SH concentrations lead to a visible darkening indicating particle clustering. We quantify these changes by means of UV-Vis spectroscopy. The bare AuNPs show a sharp absorption peak at 521 nm, characteristic of a stable dispersion of highly monodisperse particles of this size (black curve in **Fig. 2e**).[43] HMGs with 12.5 % SH also exhibit a single peak at 521 nm with no shift, confirming the absence of aggregation. Increasing SH content instead leads to a redshift in the absorption spectra[48, 49] and the emergence of a shoulder at higher wavelength with increase associated to NP clustering (**Fig. 2e**). These results highlight the importance of balancing SH-mediated AuNP binding with COOH-driven electrostatic stabilization to maintain stable HMGs. Finally, DLS measurements with temperature bath shows that MG-SH 12.5 % and HMGs have the identical temperature response and collapse around it's volume phase transition temperature (VPTT, 30-32 °C, **Fig. 2f**), indicating that the AuNPs do not strongly affect the polymer network response and architecture.

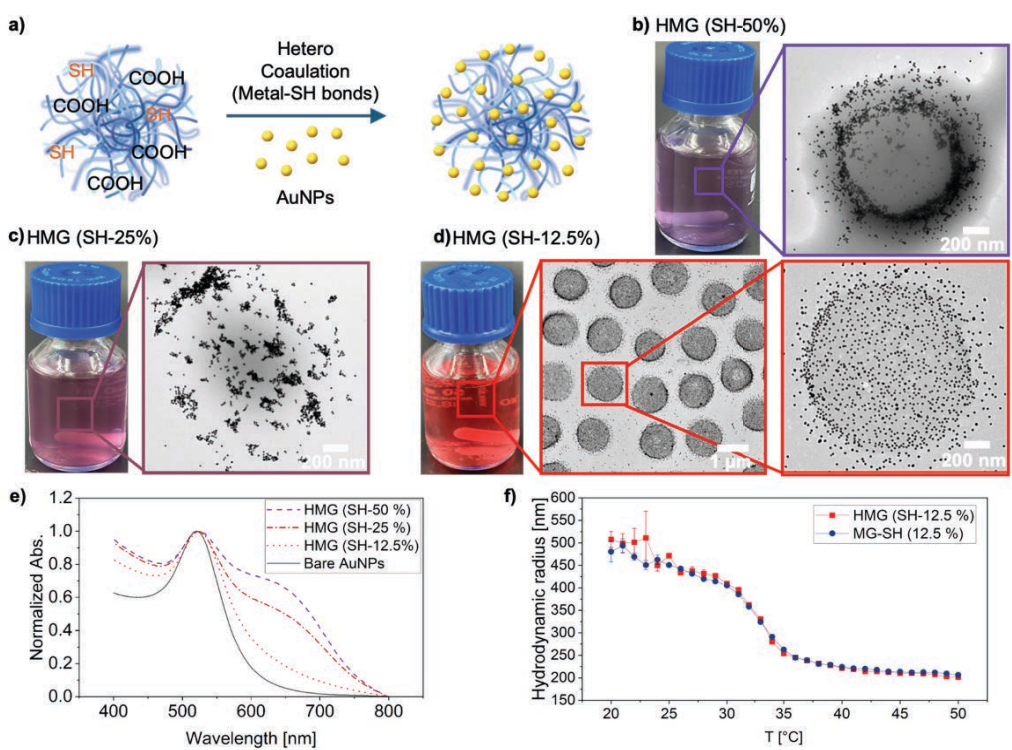

**Figure 2.** Schematic illustration of the fabrication process for HMGs (a). Photographs of HMG dispersions along with their corresponding TEM images in dry state are sown in (b)-(d). (e) UV-Vis spectra of HMGs compared with bare AuNPs. (f) T-dependent $R_h$ of HMGs and MGs with 12.5 % SH measured by DLS.

Next, we adopted the recently introduced dual-laser DLS approach developed by Lehmann et al[32] (**Fig. 3a**). In their method, a red laser (633 nm) was employed for scattering, while a green laser (532 nm) was used to induce photothermal shrinkage. Although they observed a clear size reduction of HMGs with increased green laser power, an anomalous rise in diffusivity was noted at higher scattering angles (q > 0.013 nm$^{-1}$, ~60°). They attributed this anomaly to inhomogeneous scattering and recommended limiting measurements to lower angles. Our investigation revealed that this anomalous diffusivity likely originates from an additional ballistic motion component caused by laser radiation pressure. This pressure results from the direct momentum transfer of photons to particles, leading to directional, non-random particle motion under illumination.[50, 51] Indeed, we found that autocorrelation functions of scattered intensity could be accurately described only by incorporating this ballistic motion component (**Fig. 3b**). This ballistic term has characteristic drift velocity, increasing proportionally with the green laser power (**Fig. 3c**), clearly confirming laser-induced radiation pressure as the contributing factor. By explicitly accounting for and subtracting this ballistic motion artifact, we achieved accurate hydrodynamic radius ($R_h$) measurements even at a 90° scattering angle. This improved analysis clearly confirmed that the HMGs exhibit a steady decrease in size with increasing laser intensity (**Fig. 3d**), closely matching the temperature-controlled DLS measurements (**Fig. 2d**). From these corrected data, we were able to reliably extract the local temperature changes within the HMGs as a function of green laser power density (**Fig. 3d**). Importantly, this analysis assumed that the viscosity of the surrounding medium remained constant and was unaffected by local heating, closely aligning with traditional temperature-controlled batch DLS measurement (**Fig. 2f**). For comparison, we also performed calculations assuming that the local temperature around the

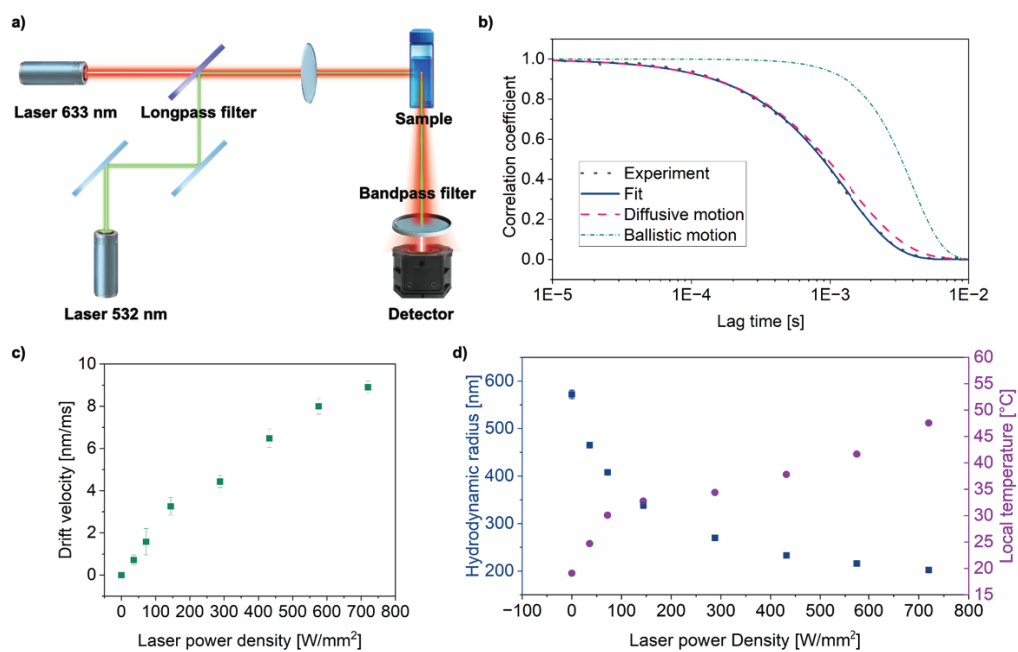

**Figure 3.** (a) Experimental setup used for photothermal responsiveness analysis using dual-laser DLS. (b) Example of scattering time autocorrelation function, displaying experimental data, pure diffusive fit, and with an additional ballistic motion component. Inclusion of ballistic motion is essential for accurately fitting the data. (c) Drift velocity extracted from the fits, showing the increase with laser power. (d) $R_h$ obtained from the fitted data as well as estimated local temperature vs laser power density. The estimated T is derived by comparing the measured $R_h$ values from T-batch DLS (**Fig. 2f**).

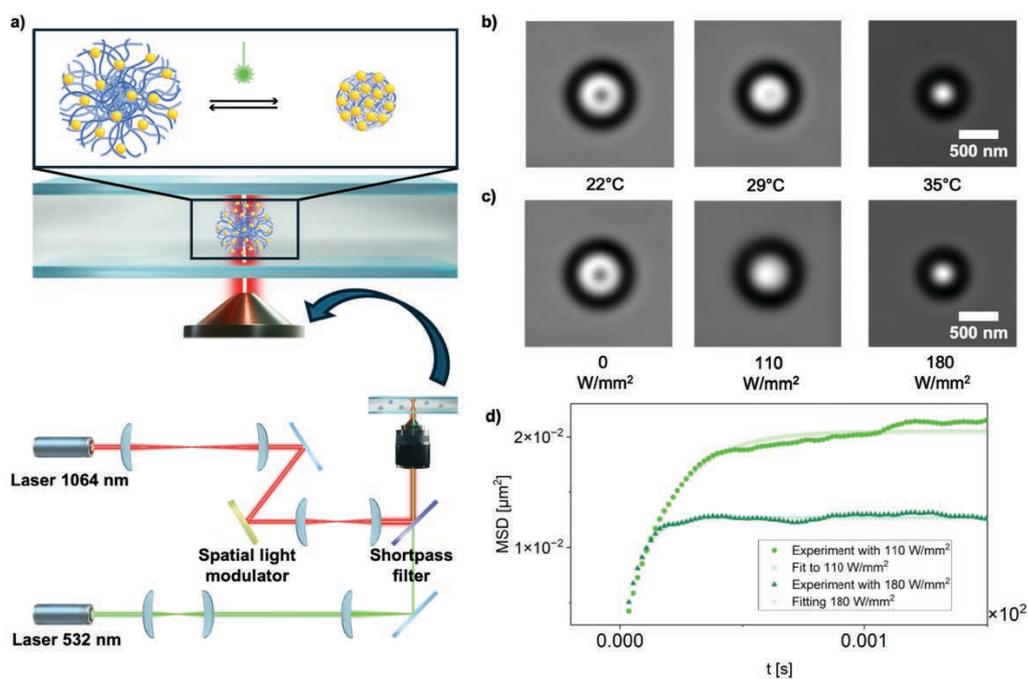

**Figure 4**. Illustration of the experimental setup for the two-laser optical microscopy (a). (b) Optical micrograph at three different temperatures as examples, and (c) optical micrographs at three different green laser powers. (d) Mean square displacement (MSD) and fitting plots at two example laser powers (110 and 180 W/mm²).

HMGs changes, thus affecting the medium's viscosity. These results are presented separately in **Figure S4**. To investigate the photo-thermal responsiveness of HMGs at the single-particle level, we developed a dual-laser optical tweezers method, where individual HMGs were trapped by an NIR laser (1064 nm) and simultaneously irradiated with a green laser (532 nm) (**Fig. 4a**). Initially, we trapped a single HMG and then gradually increased the green laser intensity until a clear collapse of the particle was visually confirmed (**Video S1**). To accurately quantify these size changes, we first established a temperature calibration by imaging the same trapped HMG at controlled temperature (without green laser irradiation). Specifically, we recorded and averaged 1000 frames per particle at each known temperature within the optical trap; representative particle images at three different temperatures are shown in **Figure 4b**. From these images, we extracted the radial pixel intensity profiles from the particle center outward, creating a characteristic calibration profile at each temperature (representative examples in **Fig. S5**). Subsequently, we repeated the imaging under green laser irradiation (plasmonic heating) using an identical imaging protocol (**Fig. 4c**). These images were then systematically compared against the full set of calibration profiles, and the temperature yielding the smallest mean residual error was selected as the estimated local temperature (representative examples in **Fig. S6**).[52] The local temperatures and corresponding radius at various initial temperatures (22°C, 25°C, and 28°C) are summarized in **Table S3**. We then established a master curve by aligning all measured HMG size at different initial temperatures to a common reference particle size of 409 nm (at 30°C, obtained via DLS measurements with T-batch, **Fig. 2f**). Each dataset was horizontally shifted along the laser power-density axis to align their HMG's sizes precisely at this reference point, explicitly setting this as zero relative power density. Negative power densities represent additional laser intensity needed to reach the 30°C reference state, while positive values indicate intensities exceeding this point. The resulting master curve (**Fig. 5a**) demonstrates a clear decrease in size with increasing relative power density. However, image analysis faced practical limitations at temperatures (and correspondingly sizes) near and above the VPTT around 30 – 32°C due to limited variations in particle contrast as a function of temperature after deswelling (an example in **Fig. S6** with 180 W/mm$^2$).

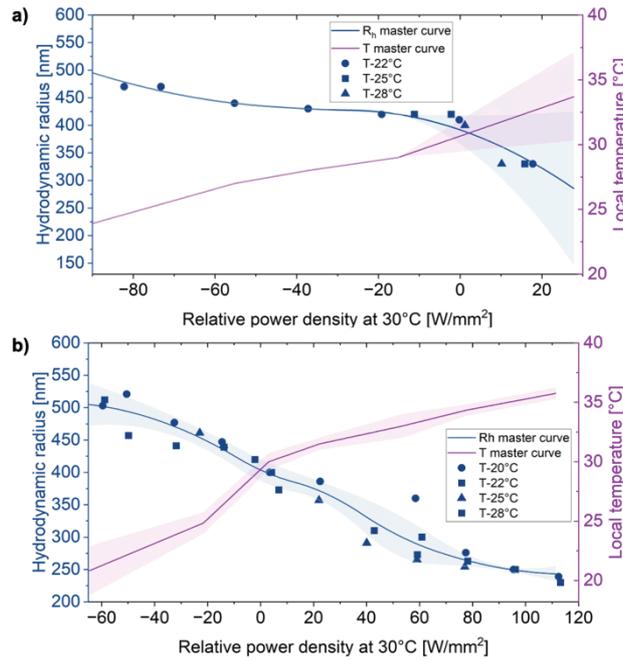

**Figure 5.** Estimation of $R_h$, radius and T using image analysis (a) and MSD-tracking based calculation (b). Solid lines represent mean values, and shaded areas indicate the error range.

To extend reliable size estimation beyond the VPTT, we extracted the particle size from mean squared displacement (MSD)-based tracking analysis using particle trajectories recorded at high frame rates (560 fps, 10,000 frames per measurement). The MSD curves were analyzed using the equation[53]:

$$<x^2(t)> = \frac{k_B T}{k}(1 - e^{-2\frac{k}{\gamma}t})$$

where $k_B$ is the Boltzmann constant, T is absolute temperature, k is the trap stiffness, and γ is the drag coefficient. From these MSD fits (example shown in **Fig. 4f**)[54], we extracted the drag coefficient γ, subsequently determining the hydrodynamic radius $R_h$ via Stokes-Einstein relation with η the medium viscosity:

$$\gamma = 6\pi\eta R$$

We chose a constant viscosity at ambient temperature for calculations, as this assumption yielded particle size data closely matching the temperature-controlled DLS results (**Fig. 3f**), suggesting minimal ambient viscosity changes around heated HMGs (raw data in **Tab. S4**). This assumption was further supported by blinking effect observed around the VPTT (30-32 °C, **Video S2**), likely reflecting dynamic equilibrium between plasmonic heating and environmental cooling.[55-57] Additional MSD analysis accounting for local viscosity changes due to localized heating are presented in **Table S5**. Similarly to the approach used in the image analysis, we normalized MSD datasets (initial temperatures of 20 °C, 23 °C, 25 °C, and 29 °C) to the common reference size of 409 nm (T = 30 °C). The resulting master curves (**Fig. 5b**) clearly gives a relation with $R_h$ (local temperature) vs relative power density. Importantly, the MSD-tracking analysis precisely captures trends even at temperatures above the VPTT, where the direct image-based size analysis becomes unreliable due to limited visual contrast.

*Conclusion*

In summary, our dual-laser optical tweezers method enables direct visualization and precise quantification of single-particle photothermal responses in HMGs. By precisely determining particle-size dynamics through image analysis and MSD-tracking calculations, we accurately quantified the local temperature induced by defined laser intensities at the single-particle level. This single-particle optical characterization approach is particularly beneficial for research fields requiring precise actuation and detailed response analysis, such as soft robotics and microswimmer design.[22, 23, 58] Moving forward, further developing this method to simultaneously measure mechanical properties alongside thermal responses, for instance through integration of stiffness via optical tweezers analysis, could substantially extend its analytical capabilities and broaden its applicability to adaptive and mechanically responsive micro-material systems.[59]

*Supporting Information*

The authors provide comprehensive details on material synthesis, calculations, experimental procedures, characterization methods, and extensive characterization results.

*Acknowledgements*

The authors thank Jose Muñeton Diaz for his support with the optical tweezers measurements, Hyeji Lee for graphical illustrations, and Ines Oberhuber from Prof. Simone Schürle's group at ETH Zurich for helping with zeta potential and DLS measurements. We also acknowledge the ScopeM facility at ETH Zurich for TEM measurements and Isabelle Feller for her valuable feedback during internal review for the manuscript preparation. Additionally, we thank Peter Kälin from the Molecular and Biomolecular Analysis Service (MoBiAS) at ETH Zurich for conducting elemental analysis. Se-Hyeong Jung and Lucio Isa acknowledges funding from the European Union's Horizon 2020 MSCA-ITN-ETN, project number 812780.

*Conflict of interest*

The authors declare no conflict of interest.

*Data Availability Statement*

The data that support the findings of this study are available from the corresponding author upon reasonable request.

**Keywords**: hybrid microgels • gold nanoparticles • optical tweezers • dynamic light scattering • Microactuator

## Table of Contents

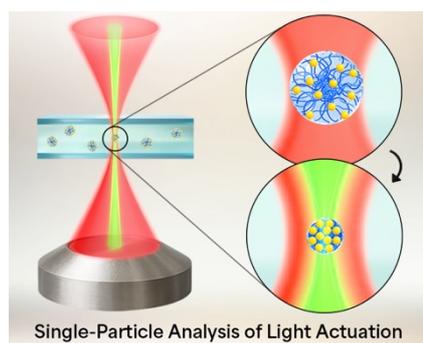

Dual-laser optical tweezers unlock real-time, single-particle resolution of light-driven actuation. Direct visualization and precise quantification of size dynamics under varying laser intensity provide insights, overcoming limitations of conventional microscale actuator analysis.

[1] Y. Jung, K. Kwon, J. Lee, and S. H. Ko, "Untethered soft actuators for soft standalone robotics," *Nature Communications,* vol. 15, no. 1, p. 3510, 2024/04/25 2024, doi: 10.1038/s41467-024-47639-0.
[2] Z. Wang, Y. Chen, Y. Ma, and J. Wang, "Bioinspired Stimuli-Responsive Materials for Soft Actuators," (in eng), *Biomimetics (Basel),* vol. 9, no. 3, Feb 21 2024, doi: 10.3390/biomimetics9030128.
[3] M. Ilami, H. Bagheri, R. Ahmed, E. O. Skowronek, and H. Marvi, "Materials, Actuators, and Sensors for Soft Bioinspired Robots," *Advanced Materials,* vol. 33, no. 19, p. 2003139, 2021, doi: https://doi.org/10.1002/adma.202003139.
[4] S. Wei and T. K. Ghosh, "Bioinspired Structures for Soft Actuators," *Advanced Materials Technologies,* vol. 7, no. 10, p. 2101521, 2022, doi: https://doi.org/10.1002/admt.202101521.
[5] M. Tyagi, J. Pan, and E. W. H. Jager, "Novel fabrication of soft microactuators with morphological computing using soft lithography," *Microsystems & Nanoengineering,* vol. 5, no. 1, p. 44, 2019/09/23 2019, doi: 10.1038/s41378-019-0092-z.
[6] W. Pang *et al.*, "A soft microrobot with highly deformable 3D actuators for climbing and transitioning complex surfaces," *Proceedings of the National Academy of Sciences,* vol. 119, no. 49, p. e2215028119, 2022, doi: doi:10.1073/pnas.2215028119.
[7] N. Vishnosky, J. C. Gomez, S. T. Kim, E. J. Doukmak, J. Grafstein, and R. C. Steinhardt, "Micron-Scale Soft Actuators Fabricated from Multi-Shell Polystyrene Particle-Gold Nanoparticle Nanohybrids," *Macromolecular Materials and Engineering,* vol. 306, no. 9, p. 2100222, 2021, doi: https://doi.org/10.1002/mame.202100222.
[8] S. Sugiura, K. Sumaru, K. Ohi, K. Hiroki, T. Takagi, and T. Kanamori, "Photoresponsive polymer gel microvalves controlled by local light irradiation," *Sensors and Actuators A: Physical,* vol. 140, no. 2, pp. 176-184, 2007/11/10/ 2007, doi: https://doi.org/10.1016/j.sna.2007.06.024.
[9] S. Palagi *et al.*, "Structured light enables biomimetic swimming and versatile locomotion of photoresponsive soft microrobots," *Nature Materials,* vol. 15, no. 6, pp. 647-653, 2016/06/01 2016, doi: 10.1038/nmat4569.
[10] T. Dinh *et al.*, "Micromachined Mechanical Resonant Sensors: From Materials, Structural Designs to Applications," *Advanced Materials Technologies,* vol. 9, no. 2, p. 2300913, 2024, doi: https://doi.org/10.1002/admt.202300913.
[11] Y. Yang *et al.*, "Microgel-Crosslinked Thermo-Responsive Hydrogel Actuators with High Mechanical Properties and Rapid Response," *Macromolecular Rapid Communications,* vol. 45, no. 8, p. 2300643, 2024, doi: https://doi.org/10.1002/marc.202300643.
[12] R. B. Yilmaz, Y. Chaabane, and V. Mansard, "Development of a Soft Actuator from Fast Swelling Macroporous PNIPAM Gels for Smart Braille Device Applications in Haptic Technology," *ACS Applied Materials & Interfaces,* vol. 15, no. 5, pp. 7340-7352, 2023/02/08 2023, doi: 10.1021/acsami.2c17835.
[13] I. U. o. P. a. A. C. (IUPAC), "Microgel," Online version 3.0.1 ed.: International Union of Pure and Applied Chemistry 2019.
[14] S.-H. Jung, S. Schneider, F. Plamper, and A. Pich, "Responsive Supramolecular Microgels with Redox-Triggered Cleavable Crosslinks," *Macromolecules,* vol. 53, no. 3, pp. 1043-1053, 2020/02/11 2020, doi: 10.1021/acs.macromol.9b01292.
[15] F. Scheffold, "Pathways and challenges towards a complete characterization of microgels," *Nature Communications,* vol. 11, no. 1, p. 4315, 2020/09/04 2020, doi: 10.1038/s41467-020-17774-5.
[16] J. Jelken *et al.*, "Tuning the Volume Phase Transition Temperature of Microgels by Light," *Advanced Functional Materials,* vol. 32, no. 2, p. 2107946, 2022, doi: https://doi.org/10.1002/adfm.202107946.
[17] S. H. Jung *et al.*, "Fabrication of pH-degradable supramacromolecular microgels with tunable size and shape via droplet-based microfluidics," *Journal of Colloid and Interface Science,* vol. 617, pp. 409-421, 2022/07/01/ 2022, doi: https://doi.org/10.1016/j.jcis.2022.02.065.
[18] S. Bulut *et al.*, "Tuning the Porosity of Dextran Microgels with Supramacromolecular Nanogels as Soft Sacrificial Templates," *Small,* vol. 19, no. 45, p. 2303783, 2023, doi: https://doi.org/10.1002/smll.202303783.
[19] S.-H. Jung *et al.*, "On-Chip Fabrication of Colloidal Suprastructures by Assembly and Supramolecular Interlinking of Microgels," *Small,* vol. 20, no. 2, p. 2303444, 2024, doi: https://doi.org/10.1002/smll.202303444.
[20] M. Rey, M. A. Fernandez-Rodriguez, M. Karg, L. Isa, and N. Vogel, "Poly-N-isopropylacrylamide Nanogels and Microgels at Fluid Interfaces," *Accounts of Chemical Research,* vol. 53, no. 2, pp. 414-424, 2020/02/18 2020, doi: 10.1021/acs.accounts.9b00528.
[21] H. F. Mathews, M. I. Pieper, S.-H. Jung, and A. Pich, "Compartmentalized Polyampholyte Microgels by Depletion Flocculation and Coacervation of Nanogels in Emulsion Droplets," *Angewandte Chemie International Edition,* vol. 62, no. 36, p. e202304908, 2023, doi: https://doi.org/10.1002/anie.202304908.
[22] L. Alvarez *et al.*, "Reconfigurable artificial microswimmers with internal feedback," *Nature Communications,* vol. 12, no. 1, p. 4762, 2021/08/06 2021, doi: 10.1038/s41467-021-25108-2.
[23] S. van Kesteren, L. Alvarez, S. Arrese-Igor, A. Alegria, and L. Isa, "Self-propelling colloids with finite state dynamics," *Proceedings of the National Academy of Sciences,* vol. 120, no. 11, p. e2213481120, 2023, doi: doi:10.1073/pnas.2213481120.
[24] Y. Xu *et al.*, "Recent Advances in Microgels: From Biomolecules to Functionality," *Small,* vol. 18, no. 34, p. 2200180, 2022, doi: https://doi.org/10.1002/smll.202200180.

[25] C. Xin et al., "Light-triggered multi-joint microactuator fabricated by two-in-one femtosecond laser writing," *Nature Communications,* vol. 14, no. 1, p. 4273, 2023/07/17 2023, doi: 10.1038/s41467-023-40038-x.

[26] Y. Li et al., "Recent Progress on Regulating the LCST of PNIPAM-Based Thermochromic Materials," *ACS Applied Polymer Materials,* vol. 7, no. 1, pp. 1-11, 2025/01/10 2025, doi: 10.1021/acsapm.4c03406.

[27] F. A. Plamper and W. Richtering, "Functional Microgels and Microgel Systems," *Accounts of Chemical Research,* vol. 50, no. 2, pp. 131-140, 2017/02/21 2017, doi: 10.1021/acs.accounts.6b00544.

[28] M. Karg and T. Hellweg, "Smart inorganic/organic hybrid microgels: Synthesis and characterisation," *Journal of Materials Chemistry,* 10.1039/B820292N vol. 19, no. 46, pp. 8714-8727, 2009, doi: 10.1039/B820292N.

[29] X. Guan, G. Cheng, Y.-P. Ho, B. P. Binks, and T. Ngai, "Light-Driven Spatiotemporal Pickering Emulsion Droplet Manipulation Enabled by Plasmonic Hybrid Microgels," *Small,* vol. 19, no. 47, p. 2304207, 2023, doi: https://doi.org/10.1002/smll.202304207.

[30] A. Sutton et al., "Photothermally triggered actuation of hybrid materials as a new platform for in vitro cell manipulation," *Nature Communications,* vol. 8, no. 1, p. 14700, 2017/03/13 2017, doi: 10.1038/ncomms14700.

[31] C. D. Jones and L. A. Lyon, "Photothermal Patterning of Microgel/Gold Nanoparticle Composite Colloidal Crystals," *Journal of the American Chemical Society,* vol. 125, no. 2, pp. 460-465, 2003/01/01 2003, doi: 10.1021/ja027431x.

[32] M. Lehmann et al., "DLS Setup for in Situ Measurements of Photoinduced Size Changes of Microgel-Based Hybrid Particles," *Langmuir,* vol. 34, no. 12, pp. 3597-3603, Mar 27 2018, doi: 10.1021/acs.langmuir.7b04298.

[33] S. Cormier, T. Ding, V. Turek, and J. J. Baumberg, "Actuating Single Nano-Oscillators with Light," *Advanced Optical Materials,* vol. 6, no. 6, p. 1701281, 2018, doi: https://doi.org/10.1002/adom.201701281.

[34] S. Cormier, T. Ding, V. Turek, and J. J. Baumberg, "Dynamic- and Light-Switchable Self-Assembled Plasmonic Metafilms," *Advanced Optical Materials,* vol. 6, no. 14, p. 1800208, 2018, doi: https://doi.org/10.1002/adom.201800208.

[35] S. Shi, L. Zhang, T. Wang, Q. Wang, Y. Gao, and N. Wang, "Poly(N-isopropylacrylamide)–Au hybrid microgels: synthesis, characterization, thermally tunable optical and catalytic properties," *Soft Matter,* 10.1039/C3SM52303A vol. 9, no. 46, pp. 10966-10970, 2013, doi: 10.1039/C3SM52303A.

[36] A. Khan, "Preparation and characterization of N-isopropylacrylamide/acrylic acid copolymer core–shell microgel particles," *Journal of Colloid and Interface Science,* vol. 313, no. 2, pp. 697-704, 2007/09/15/ 2007, doi: https://doi.org/10.1016/j.jcis.2007.05.027.

[37] A. Burmistrova, M. Richter, M. Eisele, C. Üzüm, and R. Von Klitzing, "The Effect of Co-Monomer Content on the Swelling/Shrinking and Mechanical Behaviour of Individually Adsorbed PNIPAM Microgel Particles," *Polymers,* vol. 3, no. 4, pp. 1575-1590, 2011. [Online]. Available: https://www.mdpi.com/2073-4360/3/4/1575.

[38] A. Harsányi, A. Kardos, and I. Varga, "Preparation of Amino-Functionalized Poly(N-isopropylacrylamide)-Based Microgel Particles," (in eng), *Gels,* vol. 9, no. 9, Aug 28 2023, doi: 10.3390/gels9090692.

[39] Q. Yan, H.-N. Zheng, C. Jiang, K. Li, and S.-J. Xiao, "EDC/NHS activation mechanism of polymethacrylic acid: anhydride versus NHS-ester," *RSC Advances,* 10.1039/C5RA13844B vol. 5, no. 86, pp. 69939-69947, 2015, doi: 10.1039/C5RA13844B.

[40] C. Wang, Q. Yan, H.-B. Liu, X.-H. Zhou, and S.-J. Xiao, "Different EDC/NHS Activation Mechanisms between PAA and PMAA Brushes and the Following Amidation Reactions," *Langmuir,* vol. 27, no. 19, pp. 12058-12068, 2011/10/04 2011, doi: 10.1021/la202267p.

[41] J. Turkevich, P. C. Stevenson, and J. Hillier, "A study of the nucleation and growth processes in the synthesis of colloidal gold," *Discussions of the Faraday Society,* 10.1039/DF9511100055 vol. 11, no. 0, pp. 55-75, 1951, doi: 10.1039/DF9511100055.

[42] M. Wuithschick et al., "Turkevich in New Robes: Key Questions Answered for the Most Common Gold Nanoparticle Synthesis," *ACS Nano,* vol. 9, no. 7, pp. 7052-7071, 2015/07/28 2015, doi: 10.1021/acsnano.5b01579.

[43] J. Kimling, M. Maier, B. Okenve, V. Kotaidis, H. Ballot, and A. Plech, "Turkevich Method for Gold Nanoparticle Synthesis Revisited," *The Journal of Physical Chemistry B,* vol. 110, no. 32, pp. 15700-15707, 2006/08/01 2006, doi: 10.1021/jp061667w.

[44] S. Mourdikoudis, R. M. Pallares, and N. T. K. Thanh, "Characterization techniques for nanoparticles: comparison and complementarity upon studying nanoparticle properties," *Nanoscale,* 10.1039/C8NR02278J vol. 10, no. 27, pp. 12871-12934, 2018, doi: 10.1039/C8NR02278J.

[45] T. J. Macdonald et al., "Thiol-Capped Gold Nanoparticles Swell-Encapsulated into Polyurethane as Powerful Antibacterial Surfaces Under Dark and Light Conditions," *Scientific Reports,* vol. 6, no. 1, p. 39272, 2016/12/16 2016, doi: 10.1038/srep39272.

[46] T. A. Salamone et al., "Thiol functionalised gold nanoparticles loaded with methotrexate for cancer treatment: From synthesis to in vitro studies on neuroblastoma cell lines," *Journal of Colloid and Interface Science,* vol. 649, pp. 264-278, 2023/11/01/ 2023, doi: https://doi.org/10.1016/j.jcis.2023.06.078.

[47] S. Shi, Q. Wang, T. Wang, S. Ren, Y. Gao, and N. Wang, "Thermo-, pH-, and Light-Responsive Poly(N-isopropylacrylamide-co-methacrylic acid)–Au Hybrid Microgels Prepared by the in Situ Reduction Method Based on Au-Thiol Chemistry," *The Journal of Physical Chemistry B,* vol. 118, no. 25, pp. 7177-7186, 2014/06/26 2014, doi: 10.1021/jp5027477.

[48] Y. Wang et al., "Aggregation affects optical properties and photothermal heating of gold nanospheres," *Scientific Reports,* vol. 11, no. 1, p. 898, 2021/01/13 2021, doi: 10.1038/s41598-020-79393-w.

[49] K. Saha, S. S. Agasti, C. Kim, X. Li, and V. M. Rotello, "Gold nanoparticles in chemical and biological sensing," (in eng), *Chem Rev,* vol. 112, no. 5, pp. 2739-79, May 9 2012, doi: 10.1021/cr2001178.

[50] A. Aschinger and J. Winter, "The application of dynamic light scattering to complex plasmas," *New Journal of Physics,* vol. 14, no. 9, p. 093035, 2012/09/18 2012, doi: 10.1088/1367-2630/14/9/093035.

[51] B. J. Berne and R. Pecora, *Dynamic Light Scattering: With Applications to Chemistry, Biology, and Physics*. Dover Publications, 2013.

[52] C. Zhang, J. Muñetón Díaz, A. Muster, D. R. Abujetas, L. S. Froufe-Pérez, and F. Scheffold, "Determining intrinsic potentials and validating optical binding forces between colloidal particles using optical tweezers," *Nature Communications,* vol. 15, no. 1, p. 1020, 2024/02/03 2024, doi: 10.1038/s41467-024-45162-w.

[53] M. C. Wang and G. E. Uhlenbeck, "On the Theory of the Brownian Motion II," *Reviews of Modern Physics,* vol. 17, no. 2-3, pp. 323-342, 04/01/ 1945, doi: 10.1103/RevModPhys.17.323.

[54] M. Grimm, T. Franosch, and S. Jeney, "High-resolution detection of Brownian motion for quantitative optical tweezers experiments," *Physical Review E,* vol. 86, no. 2, p. 021912, 08/13/ 2012, doi: 10.1103/PhysRevE.86.021912.

[55] N. Tang et al., "Thermal Transport in Soft PAAm Hydrogels," *Polymers,* vol. 9, no. 12, p. 688, 2017. [Online]. Available: https://www.mdpi.com/2073-4360/9/12/688.

[56] E. Yakovlev, G. Shandybina, and A. Shamova, "Modelling of the heat accumulation process during short and ultrashort pulsed laser irradiation of bone tissue," (in eng), *Biomed Opt Express,* vol. 10, no. 6, pp. 3030-3040, Jun 1 2019, doi: 10.1364/boe.10.003030.

[57] A. Mourran, H. Zhang, R. Vinokur, and M. Möller, "Soft Microrobots Employing Nonequilibrium Actuation via Plasmonic Heating," *Advanced Materials,* vol. 29, no. 2, p. 1604825, 2017, doi: https://doi.org/10.1002/adma.201604825.

[58] S. Palagi and P. Fischer, "Bioinspired microrobots," *Nature Reviews Materials,* vol. 3, no. 6, pp. 113-124, 2018/06/01 2018, doi: 10.1038/s41578-018-0016-9.

[59] M. Li, A. Pal, A. Aghakhani, A. Pena-Francesch, and M. Sitti, "Soft actuators for real-world applications," (in eng), *Nat Rev Mater,* vol. 7, pp. 235-249, Mar 2022, doi: 10.1038/s41578-021-00389-7.

[60]     J. Liu and Y. Lu, "Preparation of aptamer-linked gold nanoparticle purple aggregates for colorimetric sensing of analytes," *Nature Protocols,* vol. 1, no. 1, pp. 246-252, 2006/06/01 2006, doi: 10.1038/nprot.2006.38.

# Photo-Thermal Actuation of Hybrid Microgels with Dual Laser Optical Tweezers

Se-Hyeong Jung[a] †, Chi Zhang[b] †, Nick Stauffer[a], Frank Scheffold[b], Lucio Isa*[a]

## Materials

*N*-isopropylacrylamide (NIPAM, 97%, Sigma-Aldrich) was purified by recrystallization from hexane prior to use. Inhibitor in acrylic acid is filtered using Aluminum oxide in column before use. *N,N'*-methylenebisacrylamide (BIS, ≥99%), acrylic acid (AAc, 99%), potassium persulfate (KPS, ≥99%), 1-Ethyl-3-(3-dimethylaminopropyl)carbodiimide hydrochloride (EDC, ≥99%), N-Hydroxysuccinimide (NHS, 98%), Sodium citrate tribasic dihydrate (≥99%), Tetrachloroauric(III) acid trihydrate ($HAuCl_4 \cdot 3H_2O$, ≥99%) Hexane (≥99%), Hydrochloric acid (HCl, 37 %), Nitric acid (70%), aluminium oxide (neutral) were purchased from sigma-aldrich. Dialysis tubing (Standard xRC, 12-14 kDA MWCO) was received from Spectra/Por®. Milli-Q water was prepared using a Mili-Q ultrapure water purification system (Merk Millipore, 18.2 MΩ·cm). Cysteamin (99%) was obtained from Fluorochem Ltd. 4-(2-sulfonatoethyl)morpholin-4-ium buffer (MES buffer, 1.0 M) was purchased from thermo fisher scientific. Carbon-coated copper grids (Cu 400 mesh) were obtained from Quantifoil.

## Experiments and Synthesis

### Microgel synthesis

Microgel synthesis was performed in a 250 mL round-bottom flask. NIPAM (1.5276 g, 13.5 mmol, 90 mol%) and BIS (34.7 mg, 0.225 mmol, 1.5 mol% based on total monomer) were dissolved in 73 mL of miliQ water. AAc (103 µL, 0.108 g, 1.5 mmol, 10 mol%) was then added to this mixture under continuous stirring. The solution was degassed with nitrogen ($N_2$) for 30 min. Separately, KPS (0.035 g, 0.129 mmol) was degassed with $N_2$ for 45 min. The monomer + crosslinker mixture was gradually heated, reaching 70°C for additional 15 min. The initiator solution was subsequently added to the monomer mixture, and the reaction was stirred at 70°C for 3 h. Afterward, the reaction mixture was filtered through filter paper to remove aggregated particles, then left overnight at RT to reduce further aggregation. The purified microgels underwent three cycles of centrifugation (14,000 rcf, 20min, 20°C), with resuspension in Mili-Q water between each centrifugation step. Finally, the microgels were dialyzed against Mili-Q water for 7 days using dialysis tubing, with two changes a day of 5L miliQ water each time.

### Microgel modification with thiol

10 mL of each microgel solution was transferred into separated reaction vials, and 1 mL of 1 M MES buffer was added to each vial (resulting in approximately pH 6.85 after mixing). To activate the carboxyl groups, 77.4 mg (0.4 mmol, 2 equivalents relative to AAc) of EDC and 47.02 mg (0.4 mmol, 2 equivalents relative to AAc) of NHS were sequentially added, and the mixtures were stirred at RT for 5 min. Subsequently, cysteamine was introduced into the solutions at varying amounts: 1.95 mg for 12.5 mol%, 3.85 mg for 25 mol%, and 7.74 mg for 50 mol% modifications. The reactions were allowed to proceed overnight continuous stirring at RT. Following the reaction, each modified microgel was purified by centrifugation (14,000 rcf, 20°C, 20 min), repeated three times, with resuspension in Mili-Q water after each centrifugation step.

### Gold nanoparticle synthesis

AuNPs were synthesized following the citrate reduction method, initially developed by Turkevich[41] and later detailed by Kimling et al.[60] Before synthesis, all glassware was thoroughly cleaning using aqua regia (a freshly prepared 3:1 mixture of concentrated hydrochloric acid and nitric acid), rinsed extensively with Mili-Q water and dried with $N_2$. For synthesis, 98 mL of Mili-Q water was placed into a 250 mL two-neck flask equipped with a condenser and glass stoppers. Subsequently, 2 mL of a 50 mM (prepared by dissolving 33.9 mg $HAuCl_4 \cdot 3H_2O$ in Mili-Q water) was added, resulting in a final $HAuCl_4$ concentration of 1 mM. The solution was heated until reflux (~120°C). Then, 10 mL of a 38.8 mM sodium citrate solution (prepared by dissolving 114 mg trisodium citrate dihydrate in water) was rapidly added, and the stopper was replaced with another clean one. The reaction mixture changed color to dark red (wine-red) within 2-3 min. After 20 minutes, heating was stopped, and the mixture was allowed to cool to RT under continuous stirring. The AuNP suspension was stored at 4°C for future use.

### Fabrication of hybrid microgels

Hybrid microgels were fabricated using simple hetero-coagulation. Specifically, 0.2 mL of each microgel solution was diluted to 5 mL by adding Mili-Q water. Subsequently, 5 mL of the previously prepared AuNP solution was added to each diluted microgel solution. The mixtures were stirred overnight at RT. The following day, the mixtures were purified by centrifugation twice (8,000 rcf, 20°C, 5min each), and resuspended in Mili-Q water after each centrifugation step.

### Dynamic light scattering (DLS) with T-batch and zeta potential measurements

Single-temperature DLS and zeta potential measurements were conducted using a Litesizer 500 instrument (Anton Paar). All samples, including microgels and AuNPs, were diluted appropriately in Mili-Q water prior to analysis. DLS measurements were performed at a scattering angle of 90° and a constant temperature of 20°C. Each sample was measured three times to ensure reproducibility. Temperature-dependent DLS measurements (T-dependent curves) for hybrid microgels and microgels modified with 12.5 mol% SH groups were performed using a Nanolab 3D Dls instrument (LS instruments). Measurements were taken at a scattering angle of 90°, spanning a temperature range from 20°C to 50°C in increments of 1°C. At each temperature point, measurements were repeated three times, ensuring accurate determination of size variations as a function of temperature. The standard error of the mean (SEM) was calculated from these triplicate measurements as:

$$SEM = \frac{standard\ deviation\ (SD)}{\sqrt{n}}$$

### Transmission electron microscopy (TEM)

TEM was conducted using a JEM-1400 (JEOL) operated at an acceleration voltage of 120 kV. Samples were prepared by placing a droplet of suspension (10 µL) onto carbon-coated copper grids. The samples were allowed to air dry at RT prior to imaging. Images were captured under standard conditions to analyze and size distribution of nanoparticles and hybrid microgels.

### Atomic Force Microscopy (AFM)

Samples for AFM characterization were prepared by deposing a droplet of microgel suspension onto plasma-treated silicon wafers and allowing them to air-dry overnight in a fume hood. The measurements were conducted the following day using an Icon Dimension instrument (Bruker) operating in tapping mode under ambient conditions (air-surface). Imaging parameters included a proportional gain of 6.8, integral gain of 1.18, amplitude setpoint at 787 mV, and a scanning area of 3 µm x 3 µm. Both height and phase contrast data were recorded for detailed morphological analysis.

### UV-Vis spectroscopy

UV-Vis absorption spectra were recorded using a Cary 60 UV-Vis spectrophotometer (Agilent Technologies). Measurement performed in a wavelength range of 400-800 nm at medium scan speed at 20°C. Milli-Q water was used as a reference for background correction prior to each measurement. Samples were analyzed in standard quartz cuvettes.

**Elemental analysis (CHNS)**
Elemental analysis was performed to determine the sulfur (S) content of the microgel samples. Prior to measurement, samples were dried overnight using a freeze dryer (FreeZone 2.5 Liter, Labconco). Approximately 1 mg of each dried sample was weighed and analyzed using a HEKAtech EuroVector CHNS-O Elemental Analyzer, operated under standard combustion conditions to fully convert samples into gaseous combustion products for precise elemental quantification. The theoretical sulfur content was calculated based on the molar ratio of cysteamine incorporated during the modification step relative to the total monomer composition, assuming complete conversion and incorporation efficiency. Experimental values were compared against theoretical expectations to evaluate the efficiency of microgel functionalization.

**Dual-laser dynamic light scattering (DLS)**
A custom dual-laser DLS setup was constructed in-house (**Fig. 3a**). A green laser (532 nm, Verdi, coherent) was employed for actuation purposes, while a red laser (633 nm, He-Ne, Newport) served as the detection source. Measurements were performed at a 90° scattering angle. To prevent detector saturation, the green laser was filtered using a bandpass filter (company) before reaching the detector. Post-measurement data analysis, including fitting of autocorreclation functions (single exponential decay with drift) and extraction of decay times, was conducted using MATLAB-based routines developed in the laboratory.

**Dual-laser optical tweezers**
A custom-built dual-laser optical tweezers system was employed to trap and analyse particles while simultaneously tracking their position and shape. A loosely focused 532nm laser (Cobolt, HÜBNER Photonics) was used to induce plasmonic heating in gold-nanoparticle–loaded microgels, while bright-field illumination was provided by a white LED. Time-lapse images of the trapped particles were recorded using a high-resolution Prime 95B sCMOS camera (Teledyne Photometrics), mounted on a Nikon Ti2 microscope with an objective Apo TIRF 100X (numerical aperture of 1.49). The image sequences were analysed with custom MATLAB scripts based on the shape (Raidal intensity profile) of the particle. Mean-square displacement (MSD) analysis was performed to quantify mobility under different heating conditions, providing insight into particle dynamics and local interactions at the single-particle level.

**Supplementary Data**

| Sample | DLS at 20 °C (mean $R_h$) | Zetapotential at 20 °C |
|---|---|---|
| Microgel-COOH | 857 ± 13.28 nm | -14.5 ± 0.02 mV |
| Microgel-SH-50% | 603 ± 5.0 nm | -13.5 ± 0.75 mV |
| Microgel-SH-25% | 611 ± 19.4 nm | -13.8 ± 0.29 mV |
| Microgel-SH-12.5% | 630 ± 13.7 nm | -14.5 ± 0.28 mV |
| AuNPs | 18 ± 0.04 nm | -33 ± 2.36 mV |

**Table S1**. Mean hydrodynamic radius ($R_h$) and zeta potential values determined by DLS and electrophoretic mobility measurements, respectively.

| Sample (mol% SH) | Experimental S% (1st) | Experimental S% (2nd) | Average Experimental S% | Theoretical S% |
|---|---|---|---|---|
| Microgel-SH-50% | 1.12 | 1.22 | 1.17 | 1.45% |
| Microgel-SH-25% | 0.86 | 0.85 | 0.86 | 0.73% |
| Microgel-SH-12.5% | 0.73 | 0.82 | 0.78 | 0.36 |

**Table S2**. Sulfur content (%) determined by elemental analysis (CHNS) for microgel samples. Two experimental measurements are shown along with their average and the theoretically calculated sulfur content based on initial reagent ratios.

| Laser power density (W/mm²) | $R_h$ (nm) $T_{ini.}$-22 °C | Local T (°C) $T_{ini.}$-22 °C | $R_h$ (nm) $T_{ini.}$-25 °C | Local T (°C) $T_{ini.}$-25 °C | $R_h$ (nm) $T_{ini.}$-28 °C | Local T (°C) $T_{ini.}$-28 °C |
|---|---|---|---|---|---|---|
| 28 | 470 | 25 | 420 | 25 | 400 | 31 |
| 37 | 470 | 25 | 420 | 29 | 330 | 33 |
| 55 | 440 | 27 | 330 | 29 | ≥ $T_{VPTT}$ | ≥ $T_{VPTT}$ |
| 73 | 430 | 28 | ≥ $T_{VPTT}$ | ≥ $T_{VPTT}$ | ≥ $T_{VPTT}$ | ≥ $T_{VPTT}$ |
| 91 | 420 | 29 | ≥ $T_{VPTT}$ | ≥ $T_{VPTT}$ | ≥ $T_{VPTT}$ | ≥ $T_{VPTT}$ |
| 110 | 410 | 30 | ≥ $T_{VPTT}$ | ≥ $T_{VPTT}$ | ≥ $T_{VPTT}$ | ≥ $T_{VPTT}$ |
| 128 | 330 | 33 | ≥ $T_{VPTT}$ | ≥ $T_{VPTT}$ | ≥ $T_{VPTT}$ | ≥ $T_{VPTT}$ |

**Table S3**. Hydrodynamic radius (Rh) and local temperature estimates of HMGs, determined by dual-laser optical tweezers and image analysis. Measurements were performed at three different initial heating plate temperatures. Values above or around volume phase transition temperature (≥ $T_{VPTT}$) could not be precisely differentiated due to comparable radial intensity profiles after deswelling.

| Laser power density (W/mm²) | $R_h$ (nm) $T_{ini.}$-20 °C | Local T (°C) $T_{ini.}$-20 °C | $R_h$ (nm) $T_{ini.}$-23 °C | Local T (°C) $T_{ini.}$-23 °C | $R_h$ (nm) $T_{ini.}$-25 °C | Local T (°C) $T_{ini.}$-25 °C | $R_h$ (nm) $T_{ini.}$-29 °C | Local T (°C) $T_{ini.}$-29 °C |
|---|---|---|---|---|---|---|---|---|
| 28 | 503.0 | 20.0 | 512 | 23.0 | 461.0 | 25.0 | 420.0 | 29.0 |
| 37 | 521.0 | 20.0 | 457 | 24.0 | 442.0 | 25.0 | 373.0 | 31.0 |
| 55 | 477.0 | 23.0 | 441 | 25.0 | Blinking | Blinking | Blinking | Blinking |
| 73 | 447.0 | 25.0 | 439 | 26.0 | 357.0 | 32.0 | 310.0 | 33.0 |
| 91 | 400.0 | 30.0 | 400 | 30.0 | 291.0 | 33.0 | 300.0 | 33.0 |
| 110 | 386.0 | 31.0 | Blinking | Blinking | 265.0 | 35.0 | | |
| 128 | Blinking | Blinking | Blinking | Blinking | 254.0 | 35.0 | | |
| 146 | 360.0 | 32.0 | 273 | 34.0 | | | | |
| 165 | 276.0 | 34.0 | 263 | 35.0 | | | | |
| 183 | 250.0 | 35.0 | 250 | 35.0 | | | | |
| 200 | 239.0 | 36.0 | 230 | 37.0 | | | | |

**Table S4**. Mean-square displacement (MSD) tracking based data for HMGs, obtained by dual-laser optical tweezers analysis at four different initial temperatures. The analysis assumed that the ambient temperature was unaffected; thus, viscosity values corresponded to the respective initial temperatures. Near the volume phase transition temperature (VPTT), accurate MSD calculations were challenging due to particle blinking (**Video S2**), indicating non-equilibrium conditions arising from competing heating by the green laser and cooling by the surrounding medium. Above VPTT, stable MSD measurements and calculations were achievable.

| Laser power density (W/mm²) | $R_h$ (nm) $T_{ini.}$-20 °C | Local T (°C) $T_{ini.}$-20 °C | $R_h$ (nm) $T_{ini.}$-23 °C | Local T (°C) $T_{ini.}$-23 °C | $R_h$ (nm) $T_{ini.}$-25 °C | Local T (°C) $T_{ini.}$-25 °C | $R_h$ (nm) $T_{ini.}$-29 °C | Local T (°C) $T_{ini.}$-29 °C |
|---|---|---|---|---|---|---|---|---|
| 28 | 508 | 20 | 501 | 23 | 450 | 25 | 427 | 29 |
| 37 | 484 | 20 | 462 | 23 | 441 | 25 | 395 | 31 |
| 55 | 501 | 22 | 451 | 24 | Blinking | Blinking | Blinking | Blinking |
| 73 | 479 | 23 | 450 | 24 | 402 | 30 | 341 | 33 |
| 91 | 445 | 24 | 433 | 27 | 342 | 33 | 332 | 33 |
| 110 | 437 | 25 | Blinking | Blinking | 316 | 33 | | |
| 128 | Blinking | Blinking | Blinking | Blinking | 304 | 33 | | |
| 146 | 432 | 28 | 337 | 33 | | | | |
| 165 | 362 | 32 | 326 | 33 | | | | |
| 183 | 333 | 33 | 312 | 33 | | | | |
| 200 | 320 | 33 | 290 | 34 | | | | |

**Table S5**. Mean-square displacement (MSD) tracking based data for HMGs, obtained by dual-laser optical tweezers analysis at four different initial temperatures. The ambient temperature was considered equal to the total hybrid microgel temperature, and thus, viscosity values were adjusted accordingly. Near the volume phase transition temperature (VPTT), accurate MSD calculations were challenging due to particle blinking (**Video S2**), indicating non-equilibrium conditions arising from competing heating by the green laser and cooling by the surrounding medium. Above VPTT, stable MSD measurements and calculations were achievable.

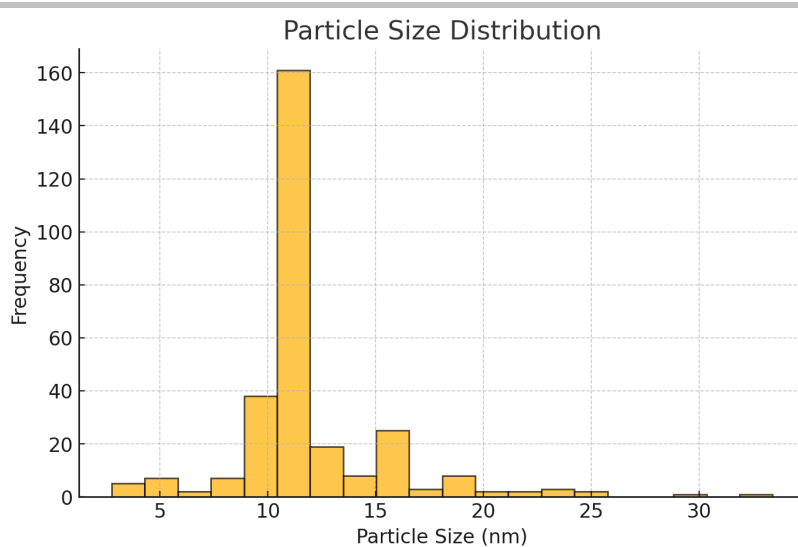

11.94 ± 0.21 nm (mean ± standard error)

**Figure S1**. Particle size distribution histogram of synthesized gold nanoparticles (AuNPs) analyzed using imageJ software. Particle diameters were extracted using Howard's method after image enhancement (contrast adjustment) and automatic thresholding Otsus method from TEM images (**Fig. 1c**).

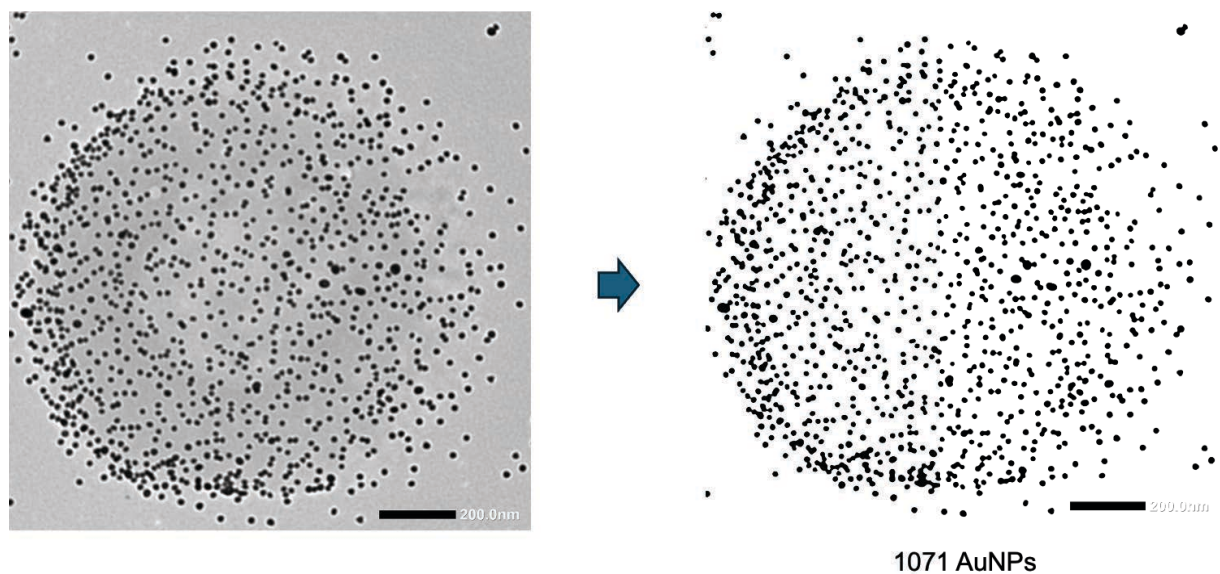

1071 AuNPs

**Figure S2**. TEM image illustrating gold nanoparticle (AuNP) localization and counting within a single hybrid microgel (**Fig. 2d**). AuNP counting was performed using ImageJ software with Howard's method after contrast enhancement and automatic thresholding (Otsu method).

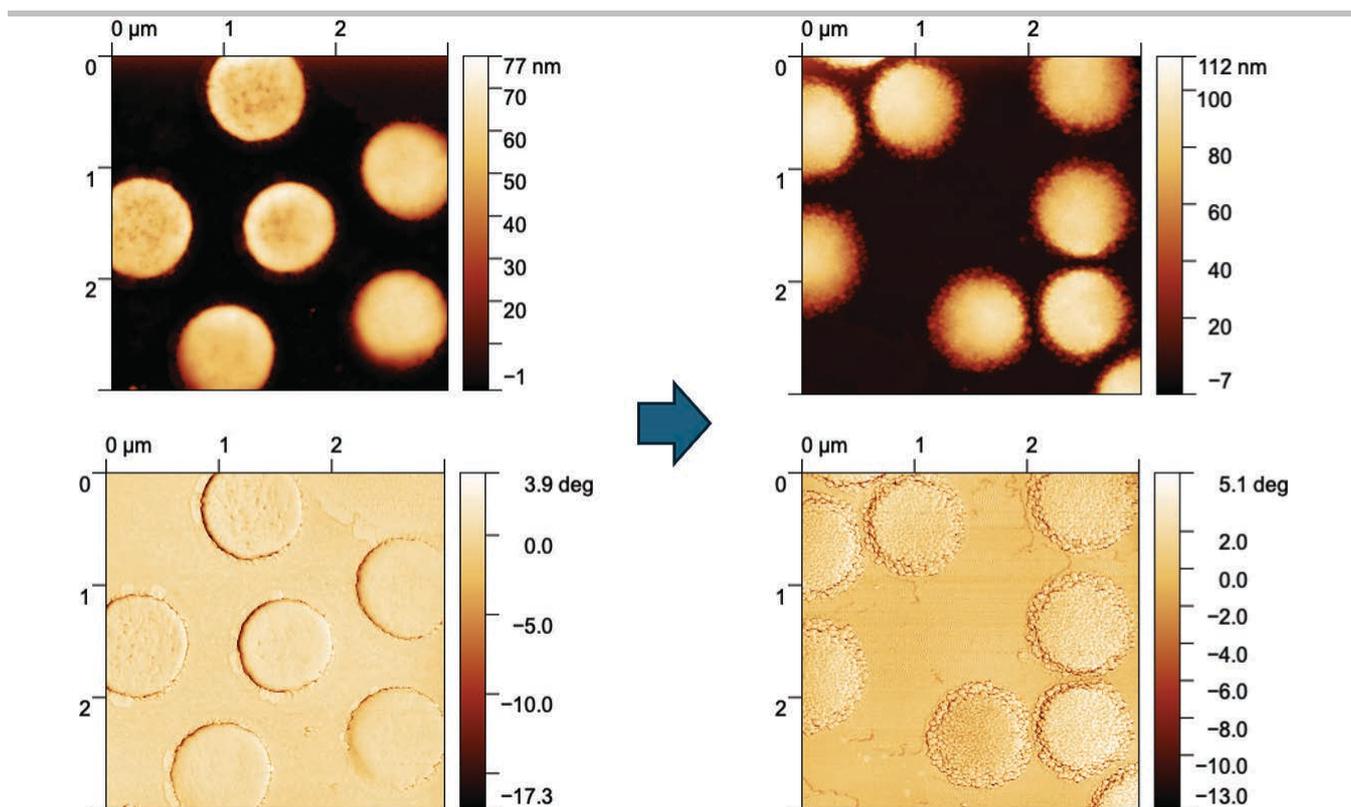

**Figure S3**. AFM height profiles (top row) and phase contrast images (bottom row) of microgels modified with 12.5 mol% SH groups (left column), and hybrid microgels (HMGs) after incorporation of AuNPs (right column). Images demonstrate morphological changes and nanoparticle distribution after AuNP integration.

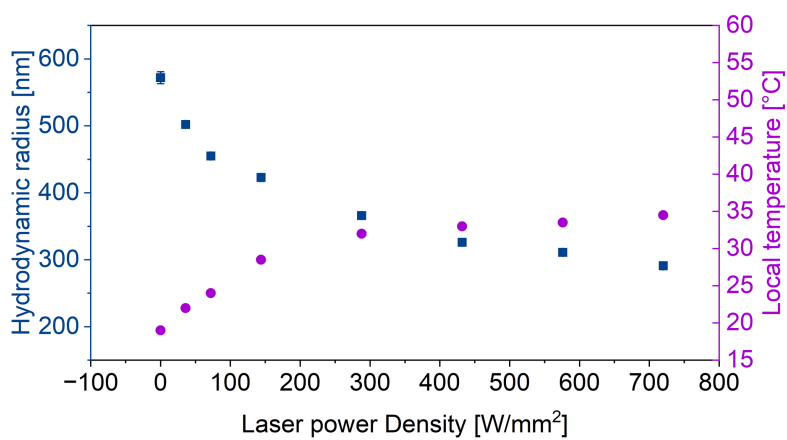

**Figure S4**. Estimation of hydrodynamic radius ($R_h$) and local temperature for hybrid microgels (HMGs) obtained from dual-laser dynamic light scattering (DLS), under the assumption that the ambient temperature and thus viscosity surrounding each HMG changes in direct accordance with the HMG's internal temperature.

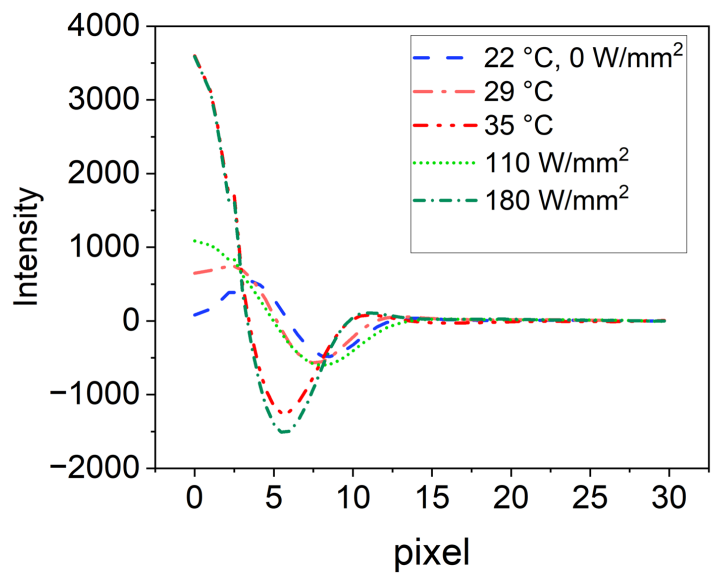

**Figure S5**. Radial intensity profiles obtained via image analysis of hybrid microgels (HMGs). Representative examples are shown at different temperatures (calibrated via T-dependent batch measurements) and varying green laser powers, illustrating how the intensity gradients evolve under these experimental conditions.

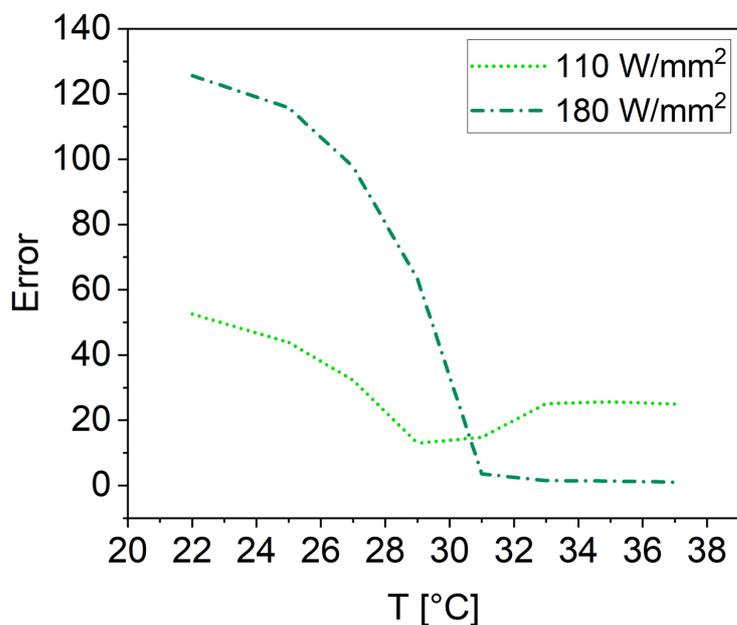

**Figure S6**. Calibration-based estimation of local temperature under plasmonic heating. Reference images of hybrid microgels (HMGs) were first acquired at different known temperatures to construct a calibration dataset. Subsequently, experimental images were recorded under two different 532,nm laser illumination intensities (110 and 180,W/mm$^2$). For each condition, the experimental image was compared to the set of calibration images, and the error was computed as a function of temperature. The estimated temperature corresponds to the calibration temperature that minimises this error. The plot shows the error versus temperature for both laser intensities. For 110,W/mm$^2$, a clear minimum is observed near T= 29 C, indicating a well-defined temperature estimate. In contrast, the 180,W/mm$^2$ condition shows relatively flat error values above the VPTT, with no distinct minimum, reflecting the fact that images above the VPTT become visually indistinguishable and thus difficult to differentiate via this image-based method.

**References**

[1]  J. Turkevich, P. C. Stevenson, J. Hillier, *Discussions of the Faraday Society* **1951**, *11*, 55-75.
[2]  J. Liu, Y. Lu, *Nature Protocols* **2006**, *1*, 246-252.